\documentclass[useAMS,usenatbib]{mn2e}
\usepackage{epsfig}
\usepackage{amsmath, amssymb,bm}
\title[Supernova Kicks and Misaligned Microquasars]{Supernova Kicks and
Misaligned Microquasars}

\author[R. G. Martin, C. A. Tout and J. E. Pringle]{Rebecca G. Martin,
  Christopher A. Tout and J. E. Pringle\\ University of Cambridge,
  Institute of Astronomy, The Observatories, Madingley Road, Cambridge
  CB3 0HA\\}
\begin{document}

\date{}

\pagerange{\pageref{firstpage}--\pageref{lastpage}}
\pubyear{2007}
\maketitle

\label{firstpage}

\begin{abstract}

  The low-mass X-ray binary microquasar GRO~J1655--40 is observed to
  have a misalignment between the jets and the binary orbital plane.
  Since the current black hole spin axis is likely to be parallel to
  the jets, this implies a misalignment between the spin axis of the
  black hole and the binary orbital plane. It is likely the black
  holes formed with an asymetric supernova which caused the orbital
  axis to misalign with the spin of the stars. We ask whether the null
  hypothesis that the supernova explosion did not affect the spin axis
  of the black hole can be ruled out by what can be deduced about the
  properties of the explosion from the known system parameters. We
  find that this null hypothesis cannot be disproved but we find that
  the most likely requirements to form the system include a small
  natal black hole kick (of a few tens of $\rm km\, s^{-1}$) and a
  relatively wide pre-supernova binary. In such cases the observed
  close binary system could have formed by tidal circularisation
  without a common envelope phase. 

\end{abstract}


\section{Introduction}

It has long been known that neutron stars have much greater space
velocities than their likely progenitors \citep{GO70} and it is now
widely accepted that that this is because they acquire large velocity
kicks in the supernova explosions in which they form \citep{S70,S78}.
The reasons for these kicks is still a matter for debate with the
leading candidates being asymmetric neutrino emission and/or asymmetric
mass release during the supernova explosion and core collapse
\citep{BP95,Pod02}.

We address here the question of the extent to which similar kicks may be
present when black holes form. Because the material forming the black holes
passes through the event horizon it is quite possible that few neutrinos
can escape \citep{gour93} so that the black hole could form with little or
no natal kick. On the other hand, asymmetric collapse to form a black hole
might lead to copious emission of gravitational waves \citep{bonnell95,
kobay03}. In addition, \cite{LL94} note that in a binary system a black
hole can form by accretion of matter on to a neutron star. In that case the
kick given to the original neutron star would appear as a kick given to the
current black hole. Evidence that black holes are indeed kicked comes from
the work of \cite{jonker04}. They looked at the out-of-plane distributions
of low mass X-ray binaries and neutron stars. They found no significant
difference between the two, leading to the conclusion that black holes are
subject to similar kicks at their formation.

In this paper we accept the evidence that stellar black holes do indeed
acquire a velocity kick when they form, but we inquire further into the
nature of the kick. In particular we ask whether the mechanism which gives
rise to the kick might also give rise to a misalignment of the black hole
spin axis with the original spin axis of the star from which it formed. The
black hole progenitor star most likely had its spin aligned with the binary
orbit before the supernova. A simple spherical collapse would preserve the
spin axis, but a more complicated collapse might not. We consider here the
simple null hypothesis that the spin axis remains unaltered by the kick
process and test the extent to which this might be contradicted by the
evidence.

We focus on the microquasar GRO J1655-40. As we
discuss in Section~\ref{GRO}, there is considerable information for this
system, about the size of the natal velocity kick and also on the
misalignment between the current black hole spin and the binary
orbital axis. In Section~\ref{model} we outline the dynamics of natal kicks
and their implications for spin/orbital misalignment and in
Section~\ref{apply} apply these results to GRO J1655-40. We discuss our
results in Section~\ref{conclude}.

\section{The microquasar GRO J1655-40}
\label{GRO}

Microquasars are black-hole X-ray binaries with relativistic radio
jets \citep{MR99}. GRO 1655-40 is a binary system consisting of a
black hole of mass $M_2 = 6.3 \pm 0.5 M_\odot$ and a lobefilling
companion star with mass $M_1 = 2.4 \pm 0.4 M_\odot$ \citep{G01}. The
binary system has a large systemic radial velocity with respect to the
Sun of $V_R =-142.4 \pm 1.5 \,\rm km\, s^{-1}$ \citep{OB97,shahbaz}
and this together with the observed proper motion led \cite{mir02} to
deduce that the system has a current space velocity of $112 \pm
18\,\rm km \, s^{-1}$.

This space velocity can be the result of a combination of two
physical causes, instantaneous mass loss during the supernova
explosion and an additional kick owing to asymmetry in the explosion
itself. The first depends on the mass lost during the
explosion as well as on the pre-supernova orbital velocity. That is,
it depends on the nature of the binary just prior to the explosion.
Further, if the amount of mass lost is too large, then an additional,
carefully directed kick may be required to keep the system bound.
Given the current position and space velocity of the orbit,
computation of the likely post-explosion orbit coupled with analysis
of the likely nature of the pre-explosion system have led
\cite{Willems05} to conclude that immediately after the formation of
the black hole the system had a space velocity in the range $45 {\rm
  \, km\, s^{-1}} < v_{\rm sys} < 115{\,\rm km\, s^{-1}} $. We make
use of this constraint in Section~\ref{apply}. They conclude that
although a symmetric supernova event (no intrinsic black hole
kick) cannot be ruled out, the constraints can be satisfied more
comfortably if the black hole did indeed have an intrinsic natal kick
of a few tens of $\rm km\, s^{-1}$. They set an upper limit to the
intrinsic kick of $210\,\rm km\,s^{-1}$.

In this paper we consider the additional constraint that the
axis of the spin of the black hole, (measured by the
direction of the relativistic jets), is misaligned with the orbital
angular velocity, along with the null hypothesis that any intrinsic
kick imparted to the black hole does not change its direction of
spin. This hypothesis implies that the kick imparts linear but
not angular momentum to the black hole.  The angle the jets make to
the line of sight was measured by \cite{H95} to be $i_{\rm
  jet}=85^\circ \pm 2^\circ$. The inclination of the binary rotation axis
to the line of sight is $i_{\rm orb} = 70^\circ.2 \pm 1^\circ.9$
\citep{OB97,G01}. So we take the angle $i$ between the black hole spin axis
and the binary orbital axis to lie in the range $ 15^\circ \pm 5^\circ \le
i \le 165^\circ \pm 5^\circ$ \citep{martin2008}. \cite{martin2008} also
show that if the black hole is spinning, interaction between the hole and
the accretion disc tends to reduce $i$. The only rigorous constraint that
we can put on the misalignment angle $i$ immediately post-supernova is that
$i > 10^\circ$ though it could be much greater than this.

The result of the mass loss coupled with any intrinsic kick would have
left the the binary system in an eccentric orbit \citep{BP95} but of
course the orbit of GRO J1655-40 is now circular.  To achieve this it
is necessary that the post-explosion orbit be such that it could be
circularised by tides.  Using the formulae given by \cite{H02} we
can estimate the circularisation timescale of the system in its
current state to be about $2\times 10^5\,$yr.  This is sufficiently
shorter than the evolution time, of at least $3\times 10^6\,$yr since
mass transfer began \citep{martin2008} and so the circular orbit now
is not an unreasonable expectation.

\section{Model}

\label{model}

In the absence of any other information on the kick it still makes sense to
see what a kick distribution like that used for neutron stars might imply
about inclinations in microquasars. We assume that prior to the supernova
the binary is in a circular orbit with two stars of masses $M_1$ and $M_2$
with {\it relative} orbital velocity $v_{\rm orb}$. We then suppose that
star 2 then has an asymmetric supernova explosion in which it loses mass
$\Delta M=M_2-M_2'$ and which gives it an intrinsic kick with velocity
magnitude $0\le v_{\rm k}<\infty$. The direction of the kick is
parameterised by the angle out of the binary plane,
$-\pi/2\le\phi\le\pi/2$, and the angle between the direction to the
instantaneous velocity of the star and the projection of the velocity kick
on to the binary orbital plane of $0\le\omega<2\pi$ (see fig.~1 of
\cite{martin09} for a diagram showing these angles). We denote the angle
between the angular momentum axes of the old and new orbits as $i$ and,
according to our null hypothesis, assume that this is the misalignment
angle between the spin of the newly formed black hole and the new binary
orbital axis. That is we assume that no angular momentum kick is imparted
to the remnant. This means that on average material ejected in the
explosion simply carries away its specific angular momentum
\citep{podsi02}.

\cite{martin09} find the angles to be related by
\begin{equation}
  \cos \omega =\frac{v_{\rm orb}}{v_{\rm k}}\frac{1}{\cos \phi}
  - \frac{|\tan \phi|}{\tan i}.
\label{re2}
\end{equation}
For a given misalignment angle, $i$, we must have $\cos \omega$ real
and so the velocity kick must lie between the locus of $\cos \omega=
1$ ($\omega=0^\circ$), which corresponds to a velocity
\begin{equation}
v_+=\frac{v_{\rm orb}}{\cos \phi}\left(1+\frac{|\tan \phi|}{\tan
i}\right)^{-1},
\label{v1}
\end{equation}
and the locus of $\cos \omega=-1$ ($\omega=180^\circ$), which
corresponds to a velocity
\begin{equation}
v_-=\frac{v_{\rm orb}}{\cos \phi}\left(-1+\frac{|\tan \phi|}{\tan
i}\right)^{-1}.
\label{v2}
\end{equation}
We consider this further in Section~\ref{apply} once we have examined the
other constraints on the kick.

\subsection{Bound Systems}

The system must remain bound after the supernova. This implies that
the velocity kick must be less than
\begin{equation}
  v_{\rm bound}=-\frac{v_{\rm orb}|\sin \phi|}{\tan i}
  + \sqrt{(3-2f) v_{\rm orb}^2+v_{\rm orb}^2 \frac{\sin^2 \phi}{\tan^2 i}}
\label{bd} \end{equation} \citep{martin09}. Here the fraction of mass lost
in the supernova from the system $f$ is given by is \begin{equation}
f=1-\frac{M'}{M}, \end{equation} where the total system mass is $M=M_1+M_2$
before the supernova and $M'=M_1+M_2'$ after it.

\subsection{System Velocity}

We now calculate the system velocity that results from mass loss coupled
with the intrinsic supernova kick. We work in the frame of the centre of
mass of the system before the supernova and follow the analysis of
\cite{BP95} and \cite{Kal96}. We use Cartesian coordinates in the centre of
mass frame before the supernova, such that the orbital plane is
perpendicular to the $z$-axis and at the time of the explosion the stars
lie on the $y$-axis. The velocities of the stars are then \begin{equation}
\bm{v}_1=\left( -\frac{M_2}{M}v_{\rm orb},0,0\right) \end{equation} and
\begin{equation} \bm{v}_2=\left( \frac{M_1}{M}v_{\rm orb},0,0\right).
\end{equation} Here $v_{\rm orb}$ is the relative orbital velocity. We
denote velocities of the stars after the explosion by $\bm{v}_1^\prime$ and
$\bm{v}_2^\prime$ respectively. After the supernova the velocity of the
companion (star 1) is unchanged so that $\bm{v}_1'=\bm{v}_1$. The velocity
of the star which experiences the supernova becomes \begin{align}
  \bm{v}_2' & = \bm{v}_2+\left(v_{\rm k} \cos \phi \cos \omega, v_{\rm
      k} \cos \phi \sin \omega,v_{\rm k} \sin \phi \right) \cr & =
  \left( v_{\rm k} \cos \phi\cos \omega + \frac{M_1}{M}v_{\rm
      orb},v_{\rm k} \cos \phi \sin \omega, v_{\rm k} \sin \phi
  \right). \end{align} The velocity of the centre of mass of the system is
zero before the supernova explosion. Afterwards it becomes \begin{align}
  \bm{v}_{\rm sys} & = \frac{M_1}{M'}\bm{v}_1+\frac{M_2'}{M'}\bm{v}_2'
  \cr & =\left( \frac{M_2'}{M'}v_{\rm k} \cos \phi \cos \omega
    -\frac{f M_1}{M'}v_{\rm orb},\right. \cr &
  \left.\frac{M_2'}{M'}v_{\rm k} \cos \phi \sin
    \omega,\frac{M_2'}{M'}v_{\rm k} \sin \phi \right).
\end{align}
The magnitude of this velocity is
\begin{align}
  v_{\rm sys}^2 = & \frac{M_2^{'2}}{M^{'2}}v_{\rm k}^2 -
  2\frac{M_2'M_1}{M^{'2}}f v_{\rm k}v_{\rm orb} \cos \phi \cos \omega
  \cr & +f^2\frac{M_1^2}{M^{'2}} v_{\rm orb}^2.
\label{vsys}
\end{align}
For a given set of masses $M_1$, $M_2$ and $M'$, the system velocity
is a function only of $v_{\rm sys}=v_{\rm sys}(\phi, i, v_{\rm k})$
because $\omega=\omega(\phi, i, v_{\rm k})$ by equation~(\ref{re2}).

\subsection{Misalignment Probability}

We are interested in the misalignment angle of the system, $i$, after
the supernova kick. This represents the angle between the old and new
angular momenta of the orbit as shown in fig.~2 of \cite{martin09}. If
$0\le i<\pi/2$ then the system is closer to alignment than
counter-alignment and if $\pi/2<i\le\pi$ then the system is closer to
counter-alignment.

We assume that the intrinsic kick is independent of the geometry of
the pre-supernova system. Thus the kick direction is taken to be
uniformly distributed over a sphere. We also assume that the velocity
distribution of the kick is a Maxwellian. Then, \cite{martin09} find
the misalignment angle probability distribution for a Maxwellian kick
distribution with velocity dispersion $\sigma_{\rm k}$ to be
\begin{equation}
P(i)di=\int_{i}^{i+di} \int \!\!\! \int_{R} I \,d\phi \,dv_{\rm k}\, di,
\label{int}
\end{equation}
where
\begin{equation}
  I=\sqrt{\frac{2}{\pi^3}}\frac{v_{\rm k}^2 }
  {2 \sigma_k^3} e^{-\frac{v_{\rm k}^2}{2\sigma_k^2}}
  \frac{|\sin \phi|}{|\sin \omega| \sin^2 i}
\label{int2}
\end{equation}
and $\omega=\omega(\phi,i, v_{\rm k})$ is defined by equation~(\ref{re2}).
The region $R$ in their figure is the region in the $\phi-v_{\rm k}$
plane for which a given misalignment angle $i$ can be produced. We
integrate this using Mathematica and Monte-Carlo methods. We consider
this further below (Section~\ref{apply}) where we apply these
particular ideas to the system GRO J1655--40.

\section{Misalignment of GRO J1655--40}
\label{apply}

In order to calculate misalignment probabilities we need to know the
properties of the system just before the supernova. In particular the
masses before and after the supernova are not known but varying them
does not significantly affect our conclusions.  So in order to have
something specific to work with we start with the state of
GRO~J1655--40 after the supernova as discussed by \cite{martin2008}.
We found two very similar models and we use the second which has
$M_1=2.8\,\rm M_\odot$ and $M_2'=5.08\,\rm M_\odot$ and the circular
period just after the supernova is $1.481 \,\rm d$ as an example. We
then assume that the progenitor was a naked helium star of mass
$M_2=10\,\rm M_\odot$. This fits with the work of \cite{Pod02},
\cite{eldridge04} and \cite{Willems05}. We expect the system to be
circular before the supernova and we need to know the relative orbital
velocity just before the explosion, $v_{\rm orb}$.  The maximum
pre-supernova circular velocity that can be achieved corresponds to
the closest binary separation that can accommodate the main-sequence
star without it filling its Roche lobe. The companion naked helium
star would be sufficiently compact to fit well inside its own lobe.
For a lobe-filling main-sequence star of mass $M_1 = 2.8\,\rm M_\odot$
and companion of mass $M_2 = 10\,\rm M_\odot$ this maximum speed is
$v_{\rm orb} = 590\,\rm km\, s^{-1}$. In practice, the probability of
forming the current system becomes very low at high values of $v_{\rm
  orb}$. We shall consider a range $50 \le v_{\rm orb}/{\rm km
  \,s^{-1}} \le 590$.

\subsection{Possible Systems}
\label{poss}

\begin{figure*}
  \epsfxsize=8.4cm \epsfbox{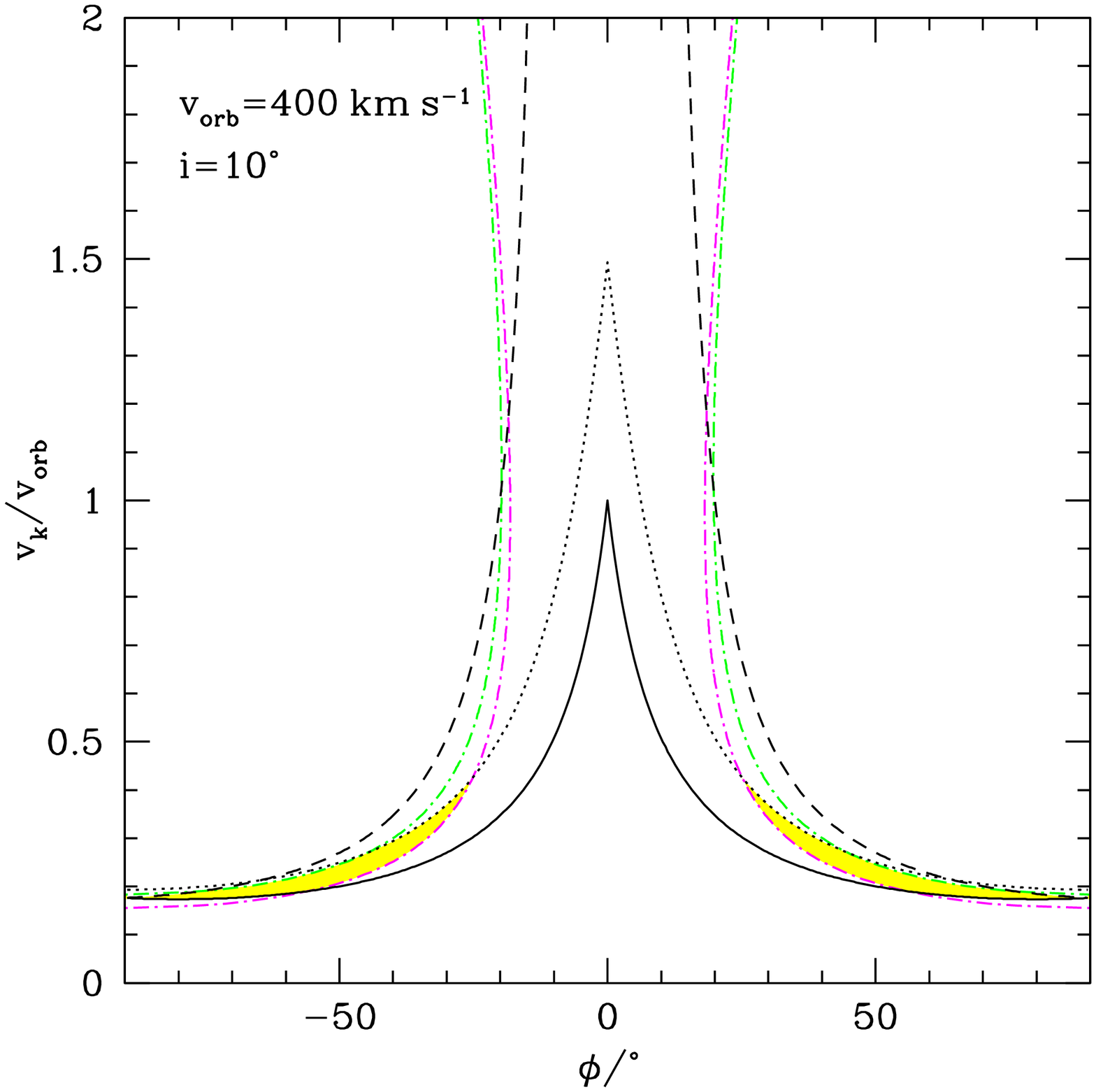} \epsfxsize=8.4cm
  \epsfbox{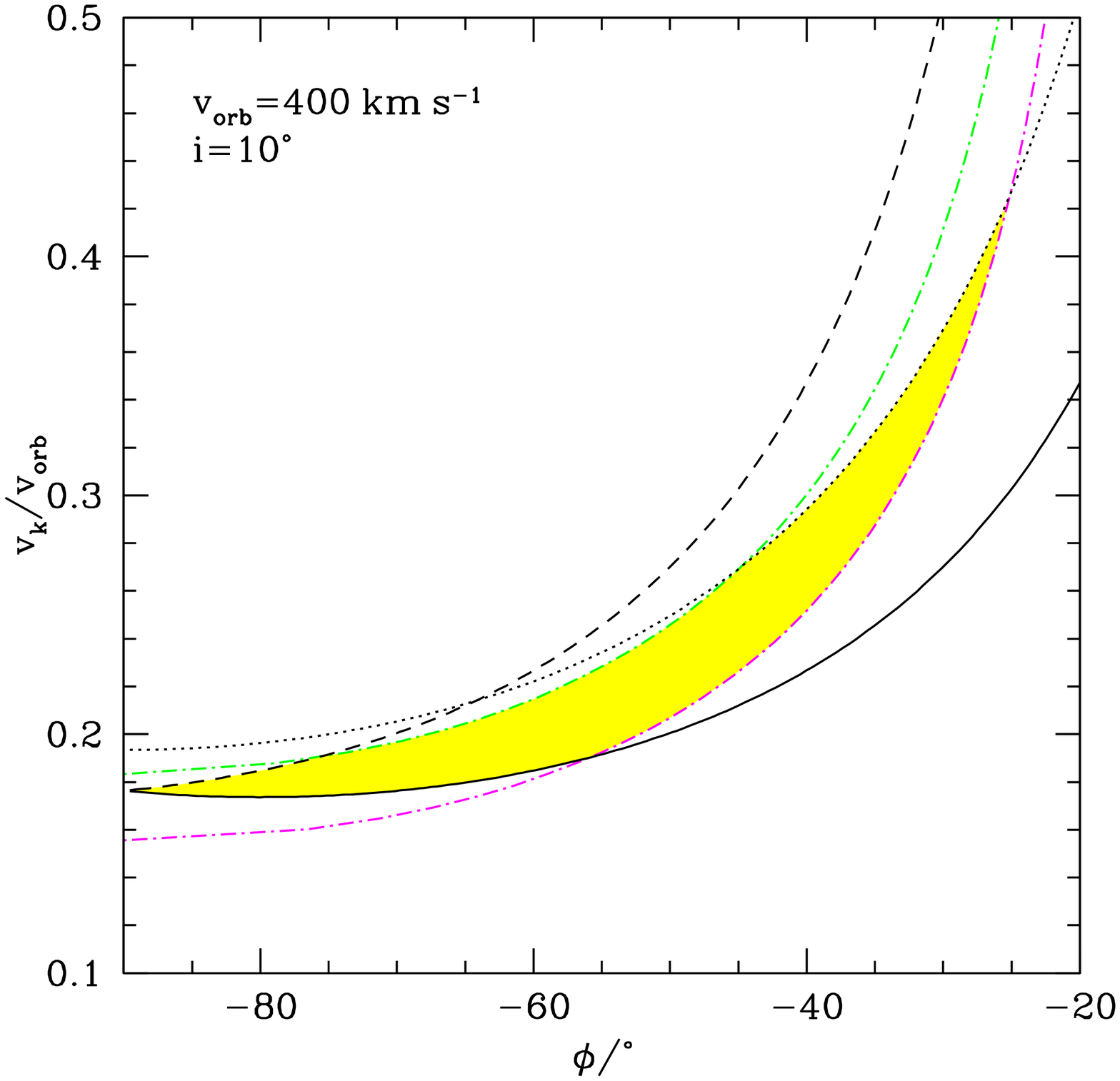} \epsfxsize=8.4cm
  \epsfbox{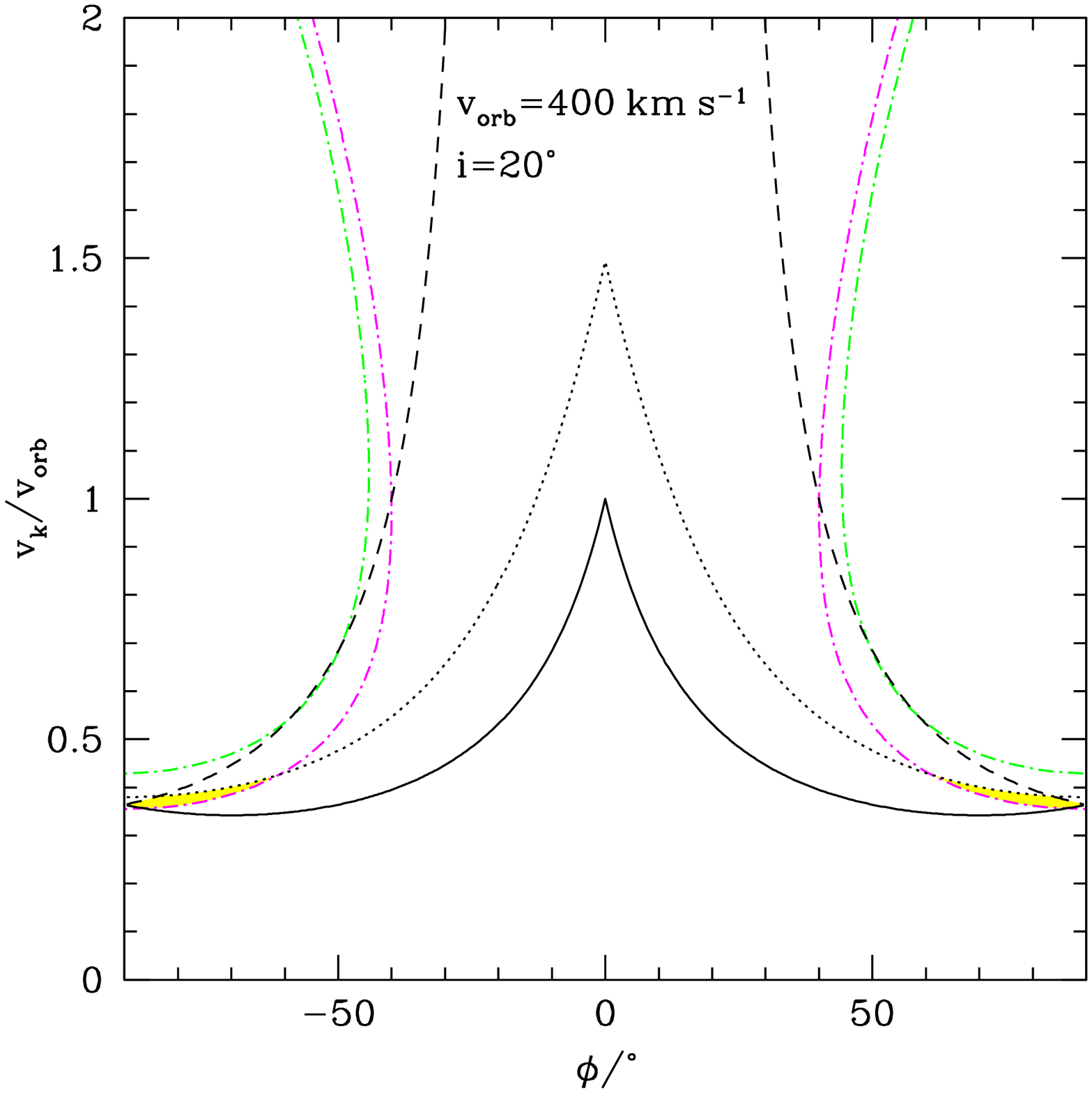} \epsfxsize=8.4cm
  \epsfbox{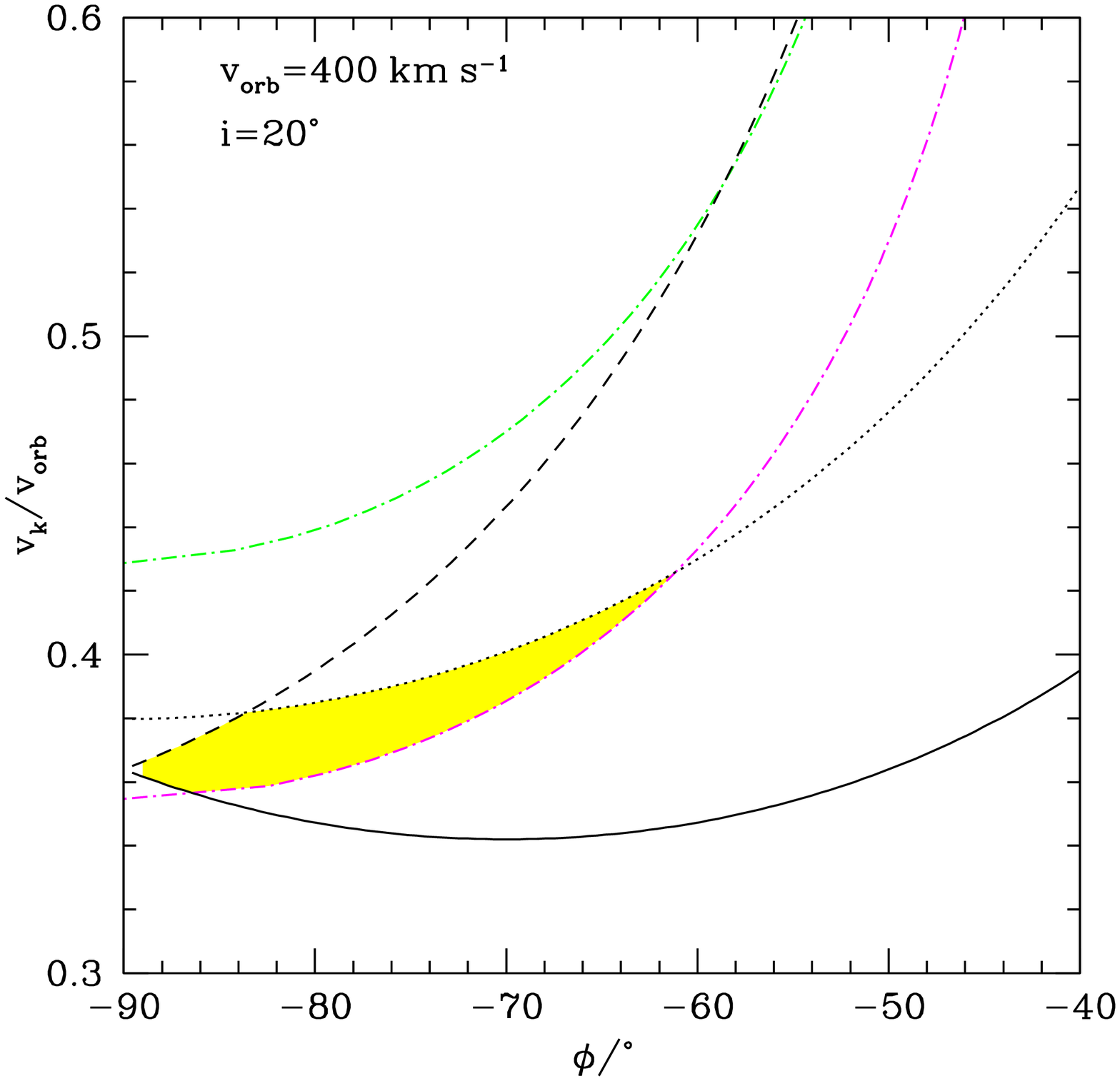}
 \caption[] {The region of the $v_{\rm k}-\phi$ plane for which the
   combination can produce a system with a misalignment of
   $i=10^\circ$ (top left and top right for a more detailed look at
   the interesting region) and $i=20^\circ$ (bottom left and bottom
   right for more detail). Here $v_{\rm k}$ is scaled by the
   pre-supernova orbital velocity $v_{\rm orb}=400\,\rm km\,s^{-1}$.
   The solid lines are the curves of $v_+$ (equation~\ref{v1}) where
   $\omega=0$.  Below the solid line $\omega$ is not real.  The dashed
   lines are $v_-$ (equation~\ref{v2}) where $\omega=180^\circ$.
   Above this line $\omega$ is not real valued.  Thus allowed values
   of $v_{\rm k}$ and $\phi$ must lie between these two curves. Below
   the dotted line the systems are bound and above they are unbound
   (equation~\ref{bd}).  Thus permitted values must also lie below the
   dotted line. In each panel the green dot-dashed line with smaller
   $\phi$ values is for $v_{\rm sys}=115\,\rm km\,s^{-1}$ and the
   other magenta one is for $v_{\rm sys}=45\,\rm km\,s^{-1}$
   (equation~\ref{vsys}).  For the post-supernova system velocity to
   lie in the required range the permitted parameters must lie between
   these two lines.  The fully constrained regions are yellow. For $i
   = 10^\circ$ the permitted range is a small region of the total
   parameter space and is even smaller for $i = 20^\circ$. }
\label{sysvel}
\end{figure*}

\begin{figure*}
  \epsfxsize=8.4cm \epsfbox{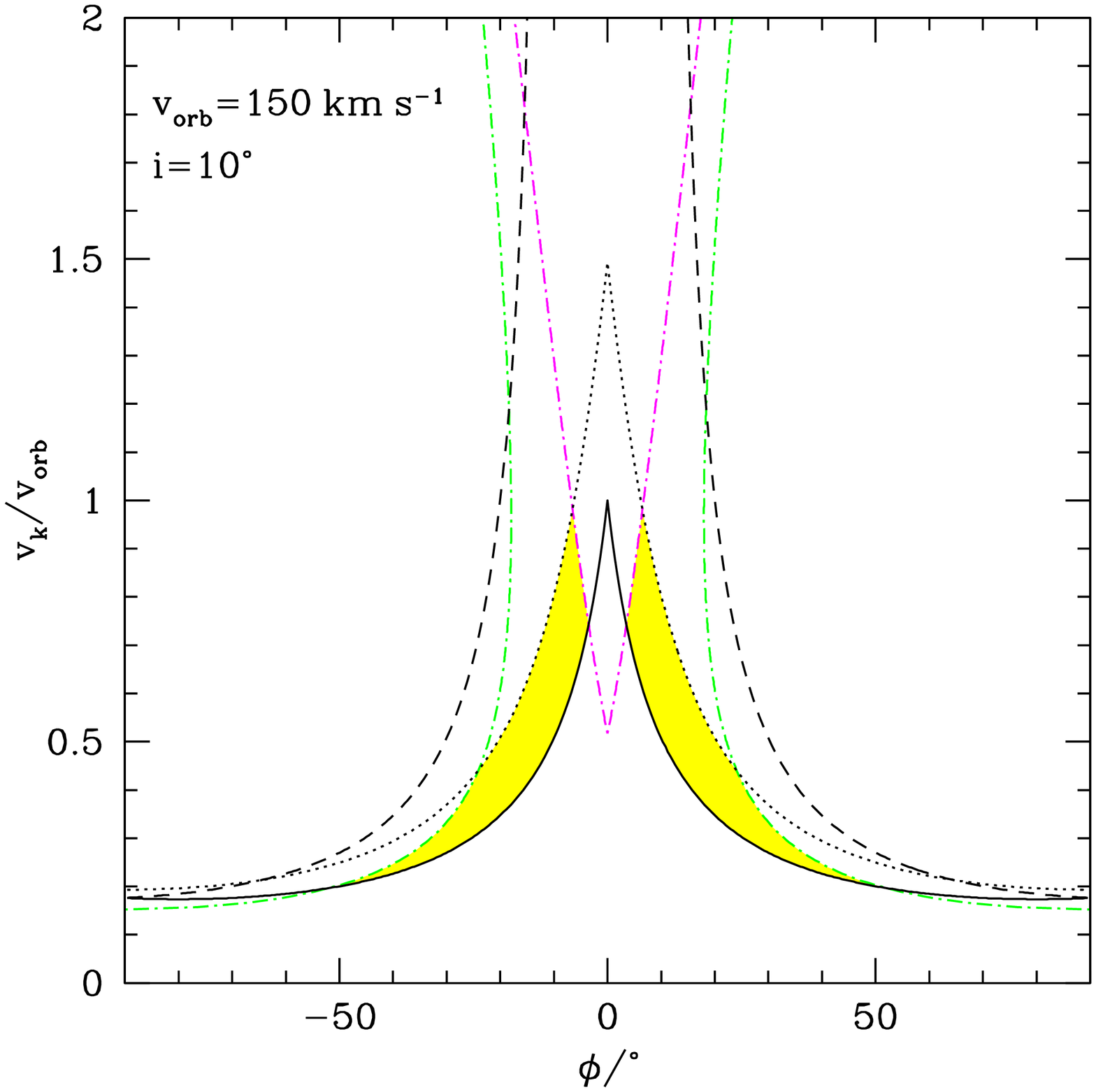} \epsfxsize=8.4cm
  \epsfbox{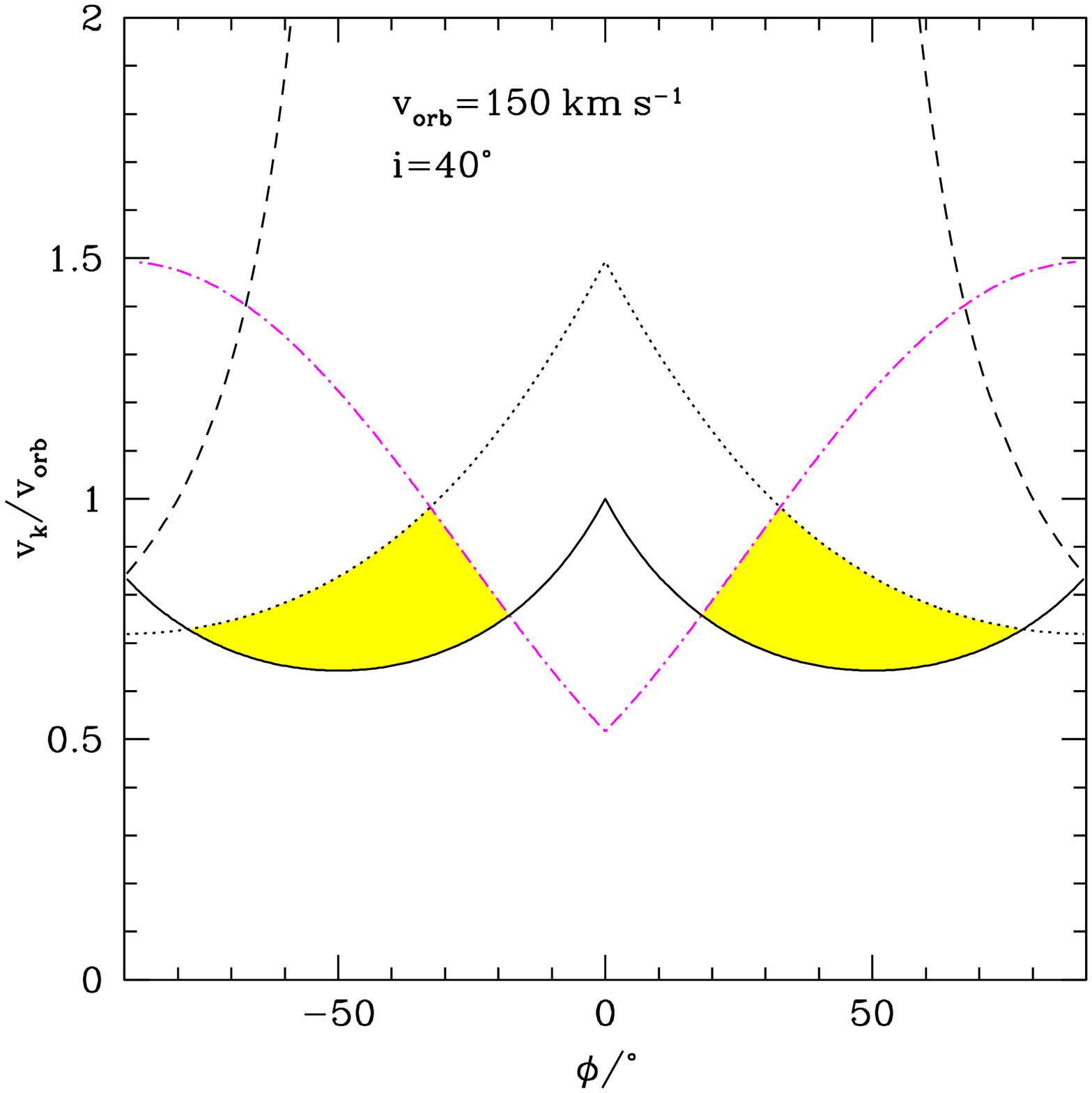}
  \caption[] {As Fig.~\ref{sysvel} but for $v_{\rm orb}=150\,\rm km\,
    s^{-1}$ with misalignments $i=10^\circ$ (left) and $i=40^\circ$
    (right). For $i = 10^\circ$ the yellow permitted region lies above
    the solid line, below the dotted line, to keep the system bound
    and below the two dot-dashed lines, to satisfy system velocity
    constraints.  For $i=40^\circ$ (right panel) the permitted region
    lies above the solid line, below the dotted line (for a bound
    system) and below the dot-dashed line to satisfy $v_{\rm sys} \le
    115\,\rm km\,s^{-1}$.  The right panel does not have a line for
    $v_{\rm sys}=45\,\rm km\,s^{-1}$ because for this no combination
    of $v_{\rm k}$ and $\phi$ can produce such a high inclination with
    such a low system velocity.  At this lower value of $v_{\rm orb}$
    the allowed regions are larger than when $v_{\rm orb}=400\,\rm
    km\,s^{-1}$.} \label{sysvel2} \end{figure*}

In Figs.~\ref{sysvel} and~\ref{sysvel2} we show, for a given
pre-explosion orbital velocity, the possible kick parameters in terms
of $v_{\rm k}$ and $\phi$ that can produce a particular misalignment
angle $i$. In Fig.~\ref{sysvel} we consider systems with
pre-supernova orbital velocity $v_{\rm orb} = 400\,\rm km\, s^{-1}$ and
post-supernova values of $i = 10^\circ$ and $i=20^\circ$. In
Fig.~\ref{sysvel2} we consider a lower value of $v_{\rm orb} = 150
\,\rm km\, s^{-1}$ and values of $i = 10^\circ$ and $i=40^\circ$. In the
$\phi-v_{\rm k}$ plane we require that the values of $\phi$ and
$v_{\rm k}$ lie in the region between the $v_+$ (equation~\ref{v1},
solid lines) and $v_-$ (equation~\ref{v2}, dashed lines) contours. In
this region we have real values of $\cos \omega$. This allowed region
is further limited by two more factors. First the system must remain
bound after the supernova. The maximum velocity kick for a system to
remain bound (equation~\ref{bd}) is shown by the dotted line.  Above
this line the system is unbound but below remains bound. Secondly the
velocity of the system after the supernova further limits the allowed
region in the $\phi-v_{\rm k}$ plane that can produce a given
inclination $i$.  We plot the lines where $v_{\rm sys}=45$ and
$115\,\rm km\,s^{-1}$ \citep{Willems05} and require that the system
must lie between these lines (equation~\ref{vsys}).

For the high value of the pre-supernova orbital velocity $v_{\rm
  orb}=400\,\rm km\,s^{-1}$ the allowed regions are very small. From
the upper panels of Fig.~\ref{sysvel} we can see that the range of
allowable values of $v_{\rm k}$ and $\phi$ is highly restricted. In
fact there are no possible combinations of $\phi$ and $v_{\rm k}$ that
can lead to a bound system with a misalignment of $i>23^\circ$ with
the required system velocity. For $v_{\rm orb}=590\,\rm km\,s^{-1}$
the space is further reduced and we cannot produce a system with
$i>17^\circ$ (see also Section~\ref{range}). A high value of $v_{\rm
  orb}$ implies that the pre-supernova system was tightly bound.
Without a kick, mass loss in the explosion results in a very unbound
system with each star having its pre-supernova velocity vector
unchanged. To keep such a system bound the kick must reduce the
post-supernova relative velocity. Such a kick must lie close to the
orbital plane and so cannot give rise to a large post-explosion
misalignment angle $i$.

For the lower value relative orbital velocity, that is for a wider
pre-supernova system, $v_{\rm orb}=150\,\rm km\,s^{-1}$, we see from
Fig.~\ref{sysvel2} that the permitted regions of parameter space are
noticeably larger although still somewhat restricted. Because the kick
no longer has to be aimed quite so accurately to cancel out the
post-explosion relative velocity it is possible to acquire larger
post-explosion misalignment angles $i$.

\subsection{Probability Distribution}

We have shown in Section~\ref{poss} that the range of values of
$v_{\rm k}$ and $\phi$ needed to give rise to the required systems is
rather restricted. And of course each individual supernova explosion
has no means of aiming for any particular restricted set. Thus, if we
assume that the kick velocities acquired by the black hole have a
certain distribution, and also that the directions of the kicks are
randomly oriented in space, we can compute a probability distribution
for the resulting misalignment angles $i$.

\begin{figure*}
  \epsfxsize=8.4cm \epsfbox{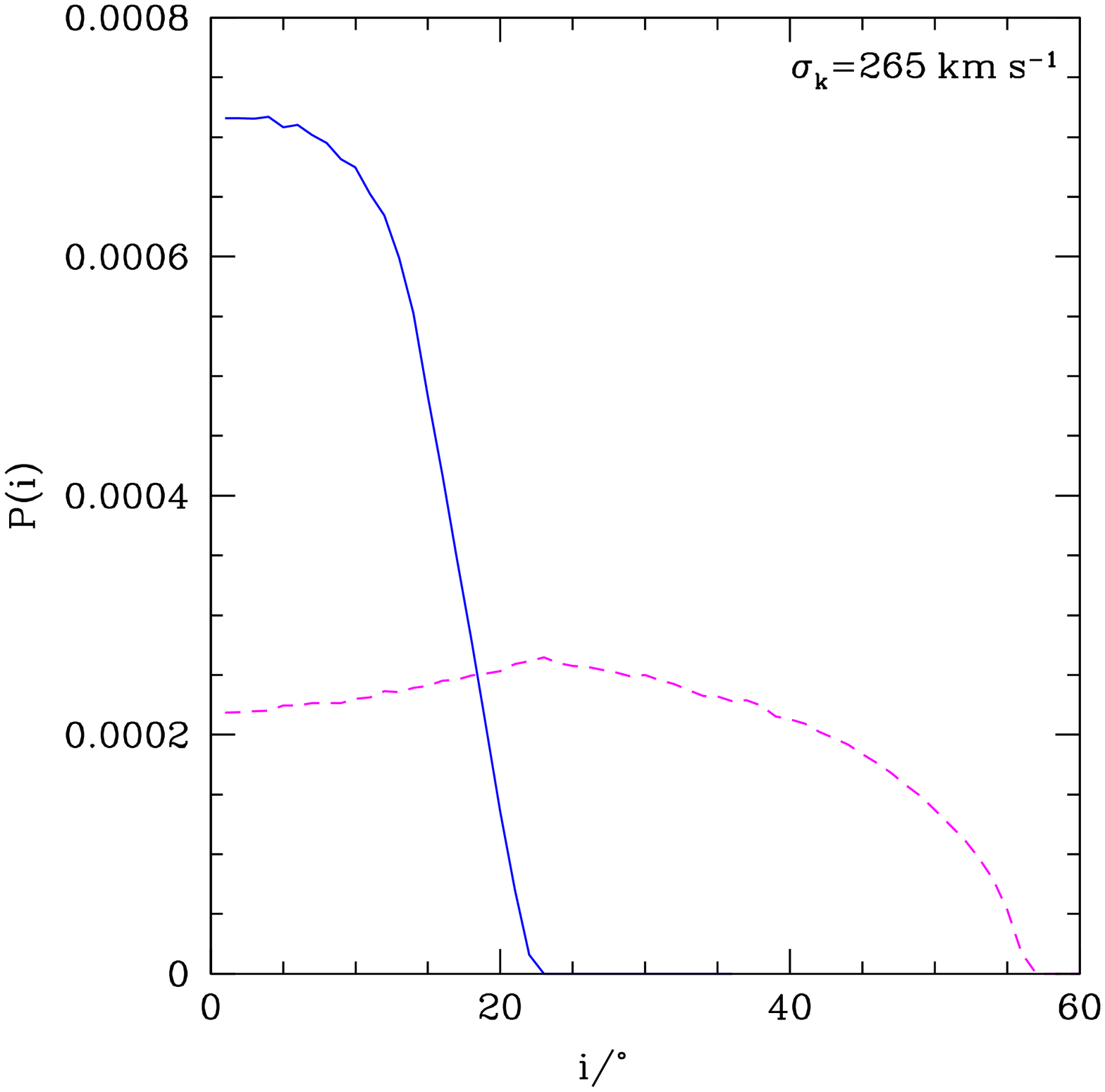} \epsfxsize=8.4cm
  \epsfbox{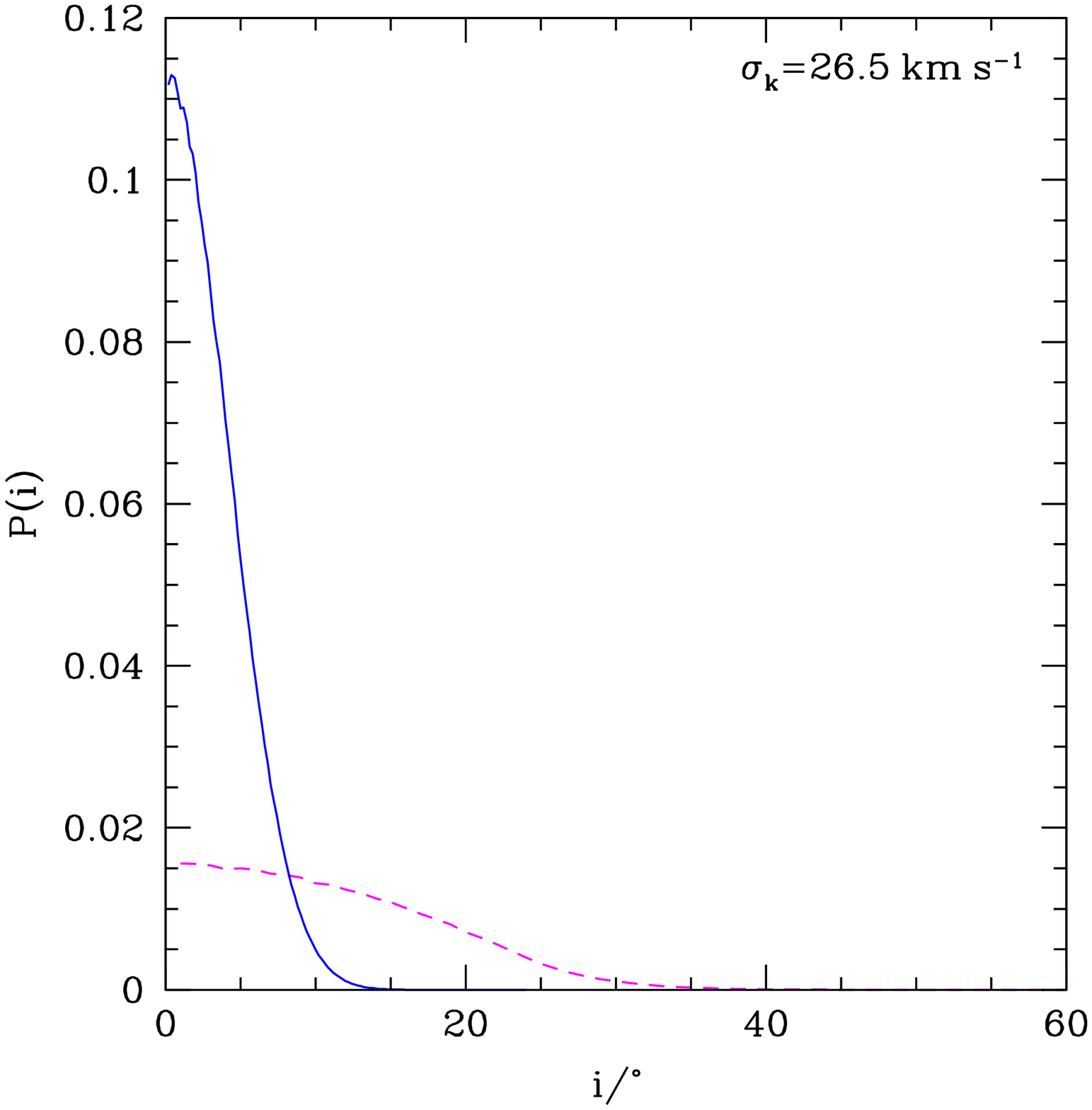}
  \caption[] {Probability distributions for the misalignment angle $i$
    between the spin of the black hole and the orbital axis for the
    microquasar GRO~J1655--40 with $v_{\rm orb}=150\,\rm km\,s^{-1}$
    (dashed lines) and $v_{\rm orb}=400\,\rm km\,s^{-1}$ (solid
    lines).  The left panel has $\sigma_{\rm k}=265\,\rm km\,s^{-1}$
    and the right has $\sigma_{\rm k}=26.5\,\rm km\,s^{-1}$. }
\label{fig:gro}
\end{figure*}

As a distribution of intrinsic kick velocities we use the standard
\cite{hobbs05} Maxwellian distribution of supernova kicks. We consider
two cases, kicks with a high velocity peak ($\sigma_{\rm k}=265\,\rm
km\,s^{-1}$) and a low velocity peak ($\sigma_{\rm k}=26.5\,\rm
km\,s^{-1}$).  For neutron stars there is some evidence that a
combination of two such velocity distributions is required to fit the
observational data \citep{arzou02}.  We integrate equation~(\ref{int})
over the yellow regions illustrated in Figs.~\ref{sysvel}
and~\ref{sysvel2} for bound systems with a system velocity in the
required range. We integrate the probability over this area using the
Monte-Carlo integrator within {\sc mathematica}. In Fig.~\ref{fig:gro}
we plot the probability distribution for two different values of the
pre-supernova orbital velocity $v_{\rm orb}$ for these two values of
$\sigma_{\rm k}$.  The high-velocity distribution of kicks (left
panel) has a much lower probability of producing the observed system
than the low-velocity distribution (right panel). Also, as we
discussed above, lower values of $v_{\rm orb}$ are more able to give
rise to larger values of $i$.

In Table~\ref{xray} we list in Column 2 the probability $P_1$ of
getting a bound system with the post-supernova system velocity in the
right range for different combinations of $\sigma_{\rm k}$ and $v_{\rm
  orb}$. In the remaining columns we show the probabilities that,
given the conditions for $P_1$ are satisfied, the misalignment angle
is greater than a particular value.  As is evident from
Fig.~\ref{fig:gro}, we see that the probability of getting a bound
system is much smaller with the higher $\sigma_{\rm k}$. However, if
we produce a bound system, then we are more likely to get higher
inclinations with the higher $\sigma_{\rm k}$. We discuss this further
in the Section~\ref{conclude}.

\begin{table*}
\begin{center}
\begin{tabular}{|r|c|c|c|c|c|c|c|c|c|}
 \hline $v_{\rm orb}/{\rm km\,s^{-1}}$ & $P_1$ & $ P(i>10^\circ)$ &$
P(i>15^\circ)$& $ P(i>20^\circ)$& $ P(i>40^\circ)$\\ \hline
 \span $\sigma_{\rm k}=265\,\rm km\,s^{-1}$ \\
\hline
 50      &       0.00155 &      0.99355 &         0.99355 &        0.98710 &
0.97419 \\
100     &       0.00967 &      0.92968 &         0.89142 &        0.85212 &
0.65460 \\
150     &       0.01179 &      0.81033 &         0.70993 &        0.60431 &
0.19161 \\
200     &       0.01209 &      0.70259 &         0.54751 &        0.39109 &
0.00088  \\
300     &       0.01174 &      0.52250 &         0.29226 &       0.11076 &
0.00000 \\
400     &       0.01143 &      0.38446 &         0.12884 &        0.00000 &
0.00000 \\
500     &       0.01136 &      0.28383 &         0.04257 &        0.00000 &
0.00000\\
\hline
 \span $\sigma_{\rm k}=26.5\,\rm km\,s^{-1}$ \\ \hline 50 & 0.06070 &
0.98575 & 0.97831 & 0.97049 & 0.93358 \\ 100 & 0.18639 & 0.73864 & 0.60720
& 0.47644 & 0.07337 \\ 150 & 0.28588 & 0.48675 & 0.27915 & 0.12740 &
0.00098\\ 200 & 0.36393 & 0.28225 & 0.09411 & 0.01917 & 0.00000 \\ 300 &
0.46989 & 0.06416 & 0.00419 & 0.00008 & 0.00000\\ 400 & 0.51754 & 0.08087 &
0.00004 & 0.00000 & 0.00000\\ 500 & 0.50958 & 0.00051 & 0.00000 & 0.00000 &
0.00000\\ \hline
 \end{tabular}
\end{center}
\caption[]{ For each combination of $\sigma_{\rm k}$ and $v_{\rm orb}$
  we show the probability that the kick produces a bound system,
  $P_1$, with the required system velocity in Column~2. In Column~3 we
  show the probability that given the conditions attached to $P_1$ are
  satisfied, it also produces a system with a misalignment greater
  than $10^\circ$ given that the system is bound and has the correct
  system velocity.  Column~4 is the same for misalignments
  $i>15^\circ$, column~5 for $i>20^\circ$ and column~6 for
  $i>40^\circ$.  To satisfy the constraints of post-supernova system
  velocity and misalignment angle, the relevant probabilities need to
  be multiplied together. }
\label{xray}
\end{table*}

\subsection{Velocity Kick Range}
\label{range}

\begin{figure}
  \epsfxsize=8.4cm \epsfbox{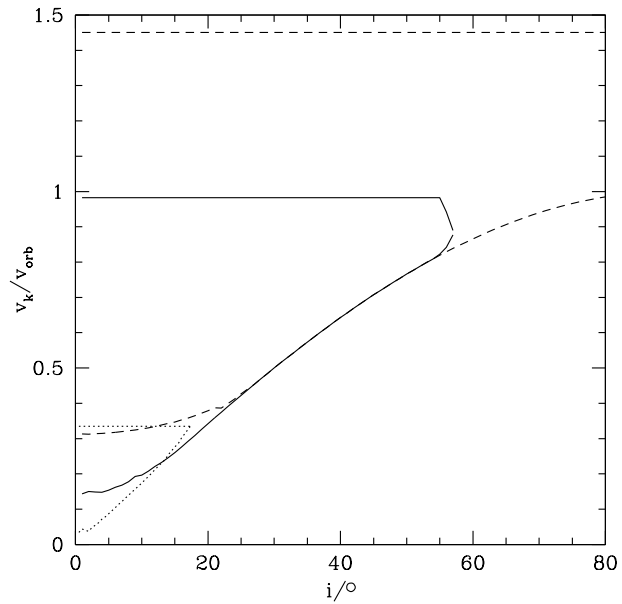}
  \caption[] {
    The range of possible velocity kicks, $v_{\rm k}$, scaled by orbital
velocity $v_{\rm orb}$ are shown for $v_{\rm orb}=100\,\rm
    km\,s^{-1}$ (dashed lines), $v_{\rm orb}=150\,\rm km\,s^{-1}$
    (solid lines) and $v_{\rm orb}=590\,\rm km\,s^{-1}$ (dotted
    lines). The upper lines show the maximum and the lower the minimum
    kick velocity that can produce a system misaligned by each angle,
    $i$.}
\label{fig:vmax}
\end{figure}

For a given initial orbital velocity, $v_{\rm orb}$, and misalignment
angle, $i$, there is a range of permissible velocity kicks, $v_{\rm
  k}$.  These ranges are illustrated for particular values of $v_{\rm
  orb}$ and $i$ with the highest and lowest velocity kick in the
permitted regions described in Figs.~\ref{sysvel} and~\ref{sysvel2}.
We plot the possible ranges of $v_{\rm k}$ as functions of $i$ for
three particular values of $v_{\rm orb}$ in Fig.~\ref{fig:vmax}. We
can deduce that the lower the pre-supernova orbital velocity the
larger the maximum possible velocity kick. To see this, we consider
the range of velocity kicks that can produce misalignments of
$i=10^\circ$, $20^\circ$ and $40^\circ$.  For $v_{\rm orb}=590\,\rm
km\,s^{-1}$ we find that we need a velocity such that $v_{\rm
  k}/v_{\rm orb} \approx 0.33$ or $v_{\rm k}\approx 198\,\rm
km\,s^{-1}$ to produce an inclination of $i=10^\circ$. This is shown
in Table~\ref{xray2}. This is the largest orbital velocity that can
produce the system so this is the largest possible velocity kick. The
ranges of permitted velocity kicks for other values of $v_{\rm orb}$
and $i$ are shown in Table~\ref{xray2}.

\begin{table*}
\begin{center}
\begin{tabular}{|r|c|c|c|c|c|c|c|c|c|}
 \hline $i/^\circ$ & $v_{\rm orb}/{\rm km\,s^{-1}}$ & $v_{\rm k min}/{\rm
km\,s^{-1}}$ & $v_{\rm k max}/{\rm km\,s^{-1}}$\\ \hline 10 & 590 & 103 &
198 \\
   & 400 & 76 & 170 \\
   & 150 & 29 & 147 \\
   & 100 & 33 & 145 \\
20 & 400 & 142 & 170 \\
   & 150  & 51 & 147 \\
   & 100 & 38 & 145 \\
40 & 150 & 96 & 147 \\
   & 100 & 64 & 145 \\
\hline
 \end{tabular}
\end{center}
\caption[]{ The range of permissible velocity kicks $v_{\rm kmin} \le
  v_{\rm k} \le v_{\rm kmax}$ for a given pre-supernova circular
  velocity $v_{\rm orb}$ that can produce the current system with
  misalignment angle $i$. These are independent of $\sigma_{\rm k}$.}
\label{xray2}
\end{table*}

\section{Conclusions}
\label{conclude}

We have considered what can be learned about the natal kick acquired
by a black hole in a supernova by considering the system GRO~1655--40.
In line with the analysis of \cite{Willems05} we make use of kinematic
data, but we add the additional constraint that the current black hole
spin is not aligned with the orbital rotation. Of course, if the
supernova explosion and collapse process in which the black hole is
formed gives not just a linear impulse, but also angular momentum, to
the hole, then the current misalignment is just a measure of the added
angular momentum. The task we set ourselves is to ask if it is
possible, given the various constraints, to rule out the possibility
that the natal kick (if any) added just linear, and not angular,
momentum. To simplify the analysis we have fixed the masses of the
stars before and after the supernova.  To be fully consistent we could
allow these masses to vary too and assign probabilities based on the
likelihood of a given set of parameters.  However one set of masses
proved to be sufficient to not rule out the null hypothesis.

Our main results are summarised in Table~\ref{xray}. We have
considered randomly oriented natal kicks with Maxwellian velocity
distributions of $265$ and $26.5\,\rm km \,s^{-1}$ for various
separations of the pre-supernova binary, parametrised in terms of its
relative orbital velocity $v_{\rm orb}$.  Column~2 gives the
probability $P_1$ that a bound system with the appropriate properties
can be formed, ignoring any information about misalignment. It is
immediately evident, that for low-velocity kicks the probability of
forming the system is high, being typically around a few tens of
percent \citep[compare][]{Willems05}. The probabilities are lower for
higher-velocity kicks, being typically around a percent. That is only
around one in a hundred such pre-supernova systems would end up
looking like GRO~1655--40. This is still probably not unreasonable.

The remaining columns in Table~\ref{xray} show the probabilities,
given $P_1$ that the additional constraint of the
misalignment angle $i$ can be satisfied, given our null hypothesis. It
is evident that we are not able to rule out this null hypothesis, and
so it is quite possible that the black hole formation process does
not affect the angular as well as the linear momentum of the
resulting black hole.

It is worth noting, however, that even without the alignment
information, it is easier to form the observed system with a
low-velocity kick.  Moreover, if the formation process did just impart
linear momentum, then, with account for the fact that the likely
alignment timescale is comparable to the age of the system since the
formation of the hole \citep{martin2008} so that the initial
misalignment angle $i$ would have been in the range $20^\circ -
40^\circ$, then it is evident that there is a strong preference for
the pre-supernova system to have be fairly wide. Note that $v_{\rm
  orb}=50\,\rm km\,s^{-1}$ corresponds to a binary separation of
around $4.6\,\rm AU$ before the supernova. It is interesting that such
a wide system could have avoided the traditional common envelope
required to shrink the orbit \citep{verbunt96}

We also note, that if the current misalignment angle were large (say,
$i > 40^\circ$) then it would become increasingly difficult to satisfy
the constraints for GRO~1655--40. In this respect we note that V4641
Sgr is a microquasar similar in many respects to GRO~1655--40 and it
has a current misalignment angle of $i \approx 55^\circ$
\citep{O01}. If further observations were able to provide a constraint
on its post-supernova systemic velocity, $v_{\rm sys}$ then it might
be able to comment more securely on the null hypotheses. For example
if it could be shown to have a large space $v_{\rm sys}$ then
satisfying the null hypothesis might be problematic.

\section*{Acknowledgements}
CAT thanks Churchill College for a Fellowship.

\label{lastpage}
\end{document}